\documentclass[useAMS,usenatbib,referee]{mn2e}
\usepackage{graphicx}  \usepackage[dvips,letterpaper]{geometry}
\usepackage{color}

\def\apj{{\rm ApJ}}
\def\mnras{{\rm MNRAS}}
\def\aap{{\rm Astronomy and Astrophysics}}

\def\simlt{\lower.5ex\hbox{$\; \buildrel < \over \sim \;$}}
\def\simgt{\lower.5ex\hbox{$\; \buildrel > \over \sim \;$}}

\def\uc05{UC05}

\newcommand{\nhplus}{\ensuremath{n_{\mathrm{H}^+}}}

\def\aasref@jnl#1{{\rm #1}}

\def\aj{\aasref@jnl{AJ}}                   \def\araa{\aasref@jnl{ARA\&A}}             \def\apj{\aasref@jnl{ApJ}}                 \def\apjl{\aasref@jnl{ApJ}}                \def\apjs{\aasref@jnl{ApJS}}               \def\ao{\aasref@jnl{Appl.~Opt.}}           \def\apss{\aasref@jnl{Ap\&SS}}             \def\aap{\aasref@jnl{A\&A}}                \def\aapr{\aasref@jnl{A\&A~Rev.}}          \def\aaps{\aasref@jnl{A\&AS}}              \def\azh{\aasref@jnl{AZh}}                 \def\baas{\aasref@jnl{BAAS}}               \def\jrasc{\aasref@jnl{JRASC}}             \def\memras{\aasref@jnl{MmRAS}}            \def\mnras{\aasref@jnl{MNRAS}}             \def\pra{\aasref@jnl{Phys.~Rev.~A}}        \def\prb{\aasref@jnl{Phys.~Rev.~B}}        \def\prc{\aasref@jnl{Phys.~Rev.~C}}        \def\prd{\aasref@jnl{Phys.~Rev.~D}}        \def\pre{\aasref@jnl{Phys.~Rev.~E}}        \def\prl{\aasref@jnl{Phys.~Rev.~Lett.}}    \def\pasp{\aasref@jnl{PASP}}               \def\pasj{\aasref@jnl{PASJ}}               \def\qjras{\aasref@jnl{QJRAS}}             \def\skytel{\aasref@jnl{S\&T}}             \def\solphys{\aasref@jnl{Sol.~Phys.}}      \def\sovast{\aasref@jnl{Soviet~Ast.}}      \def\ssr{\aasref@jnl{Space~Sci.~Rev.}}     \def\zap{\aasref@jnl{ZAp}}                 \def\nat{\aasref@jnl{Nature}}              \def\iaucirc{\aasref@jnl{IAU~Circ.}}       \def\aplett{\aasref@jnl{Astrophys.~Lett.}} \def\apspr{\aasref@jnl{Astrophys.~Space~Phys.~Res.}}
                \def\bain{\aasref@jnl{Bull.~Astron.~Inst.~Netherlands}} 
                \def\fcp{\aasref@jnl{Fund.~Cosmic~Phys.}}  \def\gca{\aasref@jnl{Geochim.~Cosmochim.~Acta}}   \def\grl{\aasref@jnl{Geophys.~Res.~Lett.}} \def\jcp{\aasref@jnl{J.~Chem.~Phys.}}      \def\jgr{\aasref@jnl{J.~Geophys.~Res.}}    \def\jqsrt{\aasref@jnl{J.~Quant.~Spec.~Radiat.~Transf.}}
                \def\memsai{\aasref@jnl{Mem.~Soc.~Astron.~Italiana}}
                \def\nphysa{\aasref@jnl{Nucl.~Phys.~A}}   \def\physrep{\aasref@jnl{Phys.~Rep.}}   \def\physscr{\aasref@jnl{Phys.~Scr}}   \def\planss{\aasref@jnl{Planet.~Space~Sci.}}   \def\procspie{\aasref@jnl{Proc.~SPIE}}

\title[Hydrogen Recombination]{
Hydrogen Recombination with Multilevel atoms}

\author[De, Baron, \& Hauschildt]{
Soma De$^{1}$, E.~Baron$^{1,2}$, P.~H.~Hauschildt$^{3}$\\
$^{1}$Homer  L. Dodge Dept.   of  Physics and Astronomy, University of Oklahoma,
Norman, OK 73019, USA\\ 
$^{2}$Computational Research Division, Lawrence Berkeley
        National Laboratory, MS 50F-1650, 1 Cyclotron Rd, Berkeley, CA
        94720 USA\\
$^{3}$Hamburger Sternwarte, Gojenbergsweg 112, 21029 Hamburg, Germany\\
}

\begin{document}

\pubyear{2010}

\maketitle

\label{firstpage}

\begin{abstract}
  Hydrogen recombination is one of the most important atomic processes
  in many astrophysical objects such as Type II supernova (SN~II)
  atmospheres, the high redshift universe during the cosmological recombination
  era, and H II regions in the  interstellar medium.  Accurate predictions of
  the ionization fraction can be quite different from those given by a
  simple solution
  if one takes into account many angular momentum sub-states,
  non-resonant processes, and calculates the rates of all atomic
  processes from the solution of the radiative transfer equation
  instead of using a Planck function under the assumption of thermal
equilibrium.  We use the general
  purpose model atmosphere code \texttt{PHOENIX} 1D to
  compare how the fundamental probabilities such as the photo-ionization
  probability, the escape probability, and the collisional de-excitation
  probability are affected by the presence of other metals in the
  environment, multiple angular momentum sub-states, and 
  non-resonant processes. Our comparisons are based on a model of SN
  1999em, a SNe Type II, 20 days after its explosion.
\end{abstract}
 

\section{Introduction}

Hydrogen recombination plays a very important role in many
astrophysical phenomena such as Type II supernovae, the interstellar medium,
and cosmic recombination.

Time-dependent recombination has been studied for the case of
cosmological recombination \citep{Zeldovich,Peebles} and for
supernovae \citep*{Chugai,dessart08,soma09}.
In supernovae this occurs when  the hydrogen recombination
time-scale becomes comparable to the age of the supernova. This effect
is found to be dominant in the early epochs of the supernova's
evolution \citep{soma09}. At later times, however, effects due to
multi-level atom
effects (the 
importance of having many angular momentum sub-states) become more
important than time-dependent phenomena. \citet{soma09}
discussed how the effective recombination timescale can
be different based on time-dependent rate equations using different
hydrogen atom models. The primary goal of \citet{soma09}
was to determine the epoch in the lifetime of a supernova (during the
photospheric phase) where 
time-dependence in the rate equations is most important. In
doing so we found  that at later times model atoms with significantly
more energy levels (that is additional
angular-momentum sub-states) have a strong effect in determining the
effective recombination time-scale.  This
issue is also important for applications other than
supernovae. In fact,  considering 
more complete atomic models is important to correctly estimate the
electron density and recover  subtle features in the
spectra. Here, we study multi-level atomic systems for
hydrogen alone using a non-LTE treatment which could alter the 
hydrogen ionization fraction and therefore produce a different temperature
structure.

Cosmological recombination codes such as RECFAST \citep{SSS99,SSS00,WMS08} and RICO \citep{FCRW09}
that deal with cosmological recombination and solve for the
free electron fraction as a function of redshift do an excellent job. Nevertheless
there are assumptions about Saha equilibrium between higher
angular momentum states.
\citet{grin_hirata10} have presented their recombination code 
RECSPARSE which is a highly detailed calculation but still uses the Sobolev
approximation for line calculations. Recently the cosmological
recombination problem has been revisted by several 
authors, paying special attention to possibly neglected atomic physics
effects. The processes studied have included two-photon decays in H~I
and He~I  \citep{DG05,WS07,CS08b,Hirata08,KI08,HS08}, 
H~I continuum absorption of the He I $\lambda 584$ line photons
\citep{SH08a,SH08b,RCS08,KIV07}, stimulated two-photon decays and two-photon
absorption \citep{Hirata08,CS06,KI08}, Raman scattering \citep{Hirata08,HS08},
a more complete set of angular momentum substates
\citep{grin_hirata10}, and forbidden transitions 
in He~I \citep{WS07,SH08a,SH08b,RCS08}. \citet{HF09} studied the
effects of partial frequency redistribution, line blanketing, and
time-dependence in Ly--$\alpha$, but even then made use of the Sobolev
approximation. Here we resolve all line profiles in the co-moving
frame and accurately treat line blanketing. We  assume complete
redistribution for the line profiles.

We focus primarily on estimating
the 
photo-ionization rate, the escape probability, and the
collisional de-excitation rate. We show how these rates
depend on the  hydrogen model  atom used and we also study the effects of the
presence of metals on the results. These different
rates ultimately lead to variations in the hydrogen
ionization fraction. We also examine
when  the 2$\gamma$ process plays the most
significant role in controlling hydrogen recombination.

In \S~\ref{sec:framework}, we outline the theoretical framework
relating the photo-ionization rate, collisional de-excitation rate,
and escape probability to the effective hydrogen recombination time.
In \S~\ref{sec:modelatoms}, we describe our approach and different
test systems that we study.  In \S~\ref{sec:results} we present our
results. In \S~\ref{sec:discuss}, we state
our conclusions and describe a possible framework for future work.

\section{Theoretical Framework for Hydrogen
  Recombination} \label{sec:framework}

The ionization fraction for hydrogen is defined as
\begin{equation}
f_{\mathrm{H}}=\frac{\nhplus}{\nhplus+n_{\mathrm{H}}}\ . 
\end{equation}
For any system the ionization fraction depends on the
net recombination rate in the system. For a
system in LTE, the ionization fraction can be calculated using the
Saha equation with knowledge of the electron density and 
temperature. On the other hand, for systems not in LTE,
it is important to solve the radiative transfer equation and the rate
equations simultaneously. The recombination rate of a system is
dependent on the number of angular momentum sub-states of the system
and also on the presence of other elements in the environment that
can contribute to the free electron density. The recombination process
in a system is determined by the net rate of photo-ionization and its
inverse process (recombination) which are modified due to line
transitions. The effectiveness (towards recombination) of a resonant
transition depends on the escape probability of the line. Also there
are non-radiative transitions that take place such as the collisional
de-excitation of electrons to lower energy levels. In addition to
these there is the downward 2$\gamma$ process that connects states
of the same parity while the
corresponding upward transition rate is very low.
Therefore it is important to study 
all these quantities to determine the correct nature of 
recombination.  We investigate the 2$\gamma$ process, the
photo-ionization rate, the escape probability and the
collisional de-excitation rates of systems with different
hydrogen atom models and metallicities of the environment. We denote
$P_{\textit{n}}$ as the photo-ionization rate from a
state with index $n$ (which is not its principal quantum
number). Similarly, we define the escape probability as $\beta_{n1}$
which is the escape probability for the resonant line photon between
the level $1s$ and the level characterized by index $n$. The collisional
de-excitation rate is defined as the rate of
de-excitation of the electron from level n into the $1s$ state. The
analytical evaluation of each of these quantities is possible if the
system is in LTE or if the system has a model atom which has just a
very few angular momentum sub-states. But for a system with many
angular momentum sub states (we consider 921 bound
levels for hydrogen) and in the presence of metals, it is necessary to
evaluate these quantities numerically from a consistent solution of
the radiative transfer equation and the rate equations.

The results presented in this paper are obtained by
  simultaneously solving the coupled radiative transfer equation
  and rate equations. The rate equations use
  multi-level atoms including a large set of line and continuum
  transitions. The upward radiative rate
  ($i\rightarrow j$) is given as $n_{i}R_{ij}$ where $i$ is a
  bound state and $j$ can be a bound  or continuum state. $n_{i}$ is
  the population for level $i$. $R_{ij}$ is given as
  \citep{mihalas78sa}:
\begin{equation}
R_{ij} = 4\pi \int_{\nu_{0}}^{\infty} \alpha_{ij} (\nu) (h\nu)^{-1} J_{\nu} d\nu \ .
\end{equation} 
Similarly the downward rate ($j \rightarrow i$) is calculated as $n_{j}(\frac{n_{i}}{n_{j}})^{\ast}R_{ji}$ where $R_{ji}$ is:

\begin{equation}
R_{ji} = 4\pi \int_{\nu_{0}}^{\infty} \alpha_{ij}(\nu)(h\nu)^{-1} \left[ (\frac{2h\nu^{3}}{c^{2}})+J_{\nu} \right] e^{-\frac{h\nu}{kT}} \ d\nu \ .
\end{equation}
 In the above equations $J_{\nu}$ is the angle averaged radiation
 intensity at frequency $\nu$. 
$T$ is the temperature, $k$ is Boltzmann's constant, $\alpha_{ij}$ is
the photo-ionization cross section for the $i \rightarrow j$
transition, and $\nu_{0}$ is the threshold frequency. The level
populations denoted with ($\ast$) are the equilibrium values as defined by
\citet{mihalas78sa}. In our calculations
the Sobolev approximation has not been used. In
order to compare our results with others we plot the values
of the Sobolev escape probability. All the level
  populations were calculated without assuming the Sobolev
  approximation.

Below we give
the equations that define $P_{n}$, $\beta_{n1}$ \citep{mihalas78sa}.
Firstly,
\begin{equation}
  P_{n}=|n_{n}R_{n\kappa}-n_{\kappa}(\frac{n_{n}}{n_{\kappa}})^{\ast}R_{\kappa
    n}|, 
\end{equation}
where $\kappa$ stands for the continuum level.

The escape probability $\beta_{n1}$ is defined using the
Sobolev escape probability as
\begin{eqnarray}
\beta_{n1} &=& \frac{1-e^{-\eta_{n1}}}{\eta_{n1}}\ ,\  \\
\eta_{n1} &=& \frac{\pi e^{2}}{m_{e}c}f_{n1}\lambda_{n1}t \left(
n_{1}-\frac{g_{1}}{g_{n}}n_{n} \right)\ ,\ 
\label{formula_beta}
\end{eqnarray}
where $f_{n1}$ is the oscillator strength of the line between
excited level n and the ground state. $n_{1}$ and $n_{n}$ are
the number densities of the ground state and the excited state
referred by index n, respectively. $\lambda_{n1}$ is the wavelength of the line,
$t$ is the time since
explosion in seconds. $g_{n}$ is the degeneracy of
  level n and $\frac{g_{1}}{g_{2}} \sim \frac{1}{n^{2}}$
\citep{mihalas78sa}. In
\S~\ref{sec:modelatoms} 
 we discuss our approach to quantify the photo-ionization
rate, the escape probability, and the collisional de-excitation
rate.

\section{Variation of hydrogen model atoms and composition}
  \label{sec:modelatoms}

\subsection{Description of the test systems}

To set up the physical structure of
each system we use the density profile and luminosity from a full NLTE
calculation of a model atmosphere in homologous expansion, with a
power-law density profile $\rho \propto v^{-7}$, 20
days after explosion, which is a reasonable fit to observed spectra of
SN~1999em at that epoch.
This  underlying density profile and the total bolometric luminosity
in the observer's frame are 
chosen to be representative of the conditions in SNe
IIP near maximum light. Give this structure we
use our general purpose model atmosphere  code
\texttt{PHOENIX} 1D to solve for the new temperature structure and level
populations under the new conditions (such as the hydrogen model atom
and the presence of other metals).  The initial model
of SN~1999em at 20 days after explosion was generated using a 31
level hydrogen model atom and solar metals were included. We call this
starting model  our base model.  Our base model was generated
treating hydrogen and other elements in NLTE. The details
of our radiative transfer code are described in \citeauthor{hbjcam99}
(\citeyear{hbjcam99,hbmathgesel04}
and references therein).
The transfer equation and the rate equations were all time independent
for the base model.
The physical properties related to our base model are given in  
Table~\ref{modeltab}.

Now we define the different systems that we experimented
upon to study the behavior of the physical quantities affecting the
recombination rate. For  convenience,  we name
the different systems in the following way. We refer to Model A as
the system which has only hydrogen and the hydrogen model atom in this
case has just 4 bound states ($1s$, $2s$,
$2p_{\frac{1}{2}}$,$2p_{\frac{3}{2}}$) and 4 lines that couple those
states.   Model C is the same as  Model A, except it has a solar
composition of  metals
present in the system in addition to hydrogen and all the other metals
are treated in LTE.  Model B has only hydrogen in the system but the
hydrogen model atom has 921 bound levels and 995 lines that couple the
bound states.  Model D is the same as  Model B except this has 
metals present  and the metals are treated in LTE as
in  Model C.   The LTE treatment of  metals was
chosen to reduce computing time and in a hydrogen rich environment the
main role that metals play in continuum processes is to be a
source of additional electrons. We did not focus on how their atomic
structure affects hydrogen recombination. 

For  Models A and C, there are primarily three major
processes considered that couple the bound states:
\begin{eqnarray}
\ H_{1s} +\gamma &\rightarrow& e^{-} +H^{+}\ ;\ \nonumber \\
H_{1s}+\gamma  &\leftrightarrow& H_{2p}\ ;\  \nonumber \\
H_{2s}&\rightarrow& H_{1s}+ 2\gamma . \nonumber
\end{eqnarray}
In addition to this the $1s$ and $2p$ states are coupled by
spontaneous transitions.  In  models B and D the atomic line data were
taken from \cite{Kurucz}.  
\section{Results }\label{sec:results}

We study the rates and the resulting  hydrogen
ionization fraction, $f_{H}$, among the four different systems that have
been described in \S~\ref{sec:modelatoms}. These different systems can have
different ionization fractions primarily because the
solution of the transfer equation and the resulting temperature
structure will be different than 
 in the base model.  The models
that have metals  (Models C and D) are expected to  have
temperature structures similar to the base model. The 
4-level hydrogen atom model without  metals is
likely to have a different temperature structure than the base
model. These differences could be subtle but important enough to 
change to the hydrogen ionization fraction. In our results we
have also studied whether the 2$\gamma$ process has a significant effect on
the recombination time scale or the resulting ionization level or the
consequent temperature structure of the system.

We define
$\tau_{std}$ as the optical depth corresponding to the continuum
opacity at 500~nm.  We classify our results depending
on whether the
temperature structure was held fixed at the base model value or was
allowed to reach 
radiative equilibrium under the particular assumptions of each case, in
order to isolate the effect of 
temperature change on the level populations and the hydrogen ionization
fraction.

\begin{table*}
\centering
\caption[modeltab]{\label{modeltab}
Physical characteristics from our base model.}
\begin{tabular}{rrrrrrr}
\hline&\\
  $\tau$ & $\rho$ (g cm$^{-3}$) & R (cm) & $v$ (km s$^{-1}$) &
  $T$ (K) & $n_{e}$ (cm$^{-3}$) & $\mu$ \\ 
\hline & \\
  9.897E-05 & 1.823E-15 & 2.454E+15 & 14200.00 &   3157.61 &   99464.3 & 1.34 \\
  3.626  & 3.084E-13  & 1.242E+15 &   7187.50  &  7345.20 & 1.079E+11  & 0.75 \\
  804.480  & 6.533E-12 & 4.104E+14 &   2375.00 &    26155.8 &  2.97E+12 & 0.67 \\
\hline
\end{tabular}

 \medskip
The columns give various quantities as a function of optical
depth. $\rho$ is the mass density, $R$ is the radius, $v$ is the
velocity, $T$ is the matter temperature, $n_e$ is the electron
density, and $\mu$ is the mean molecular weight.
\end{table*}

\subsection{Metal-rich Models}

\begin{figure*}
\centering
\includegraphics[width=0.65\textwidth,angle=0]{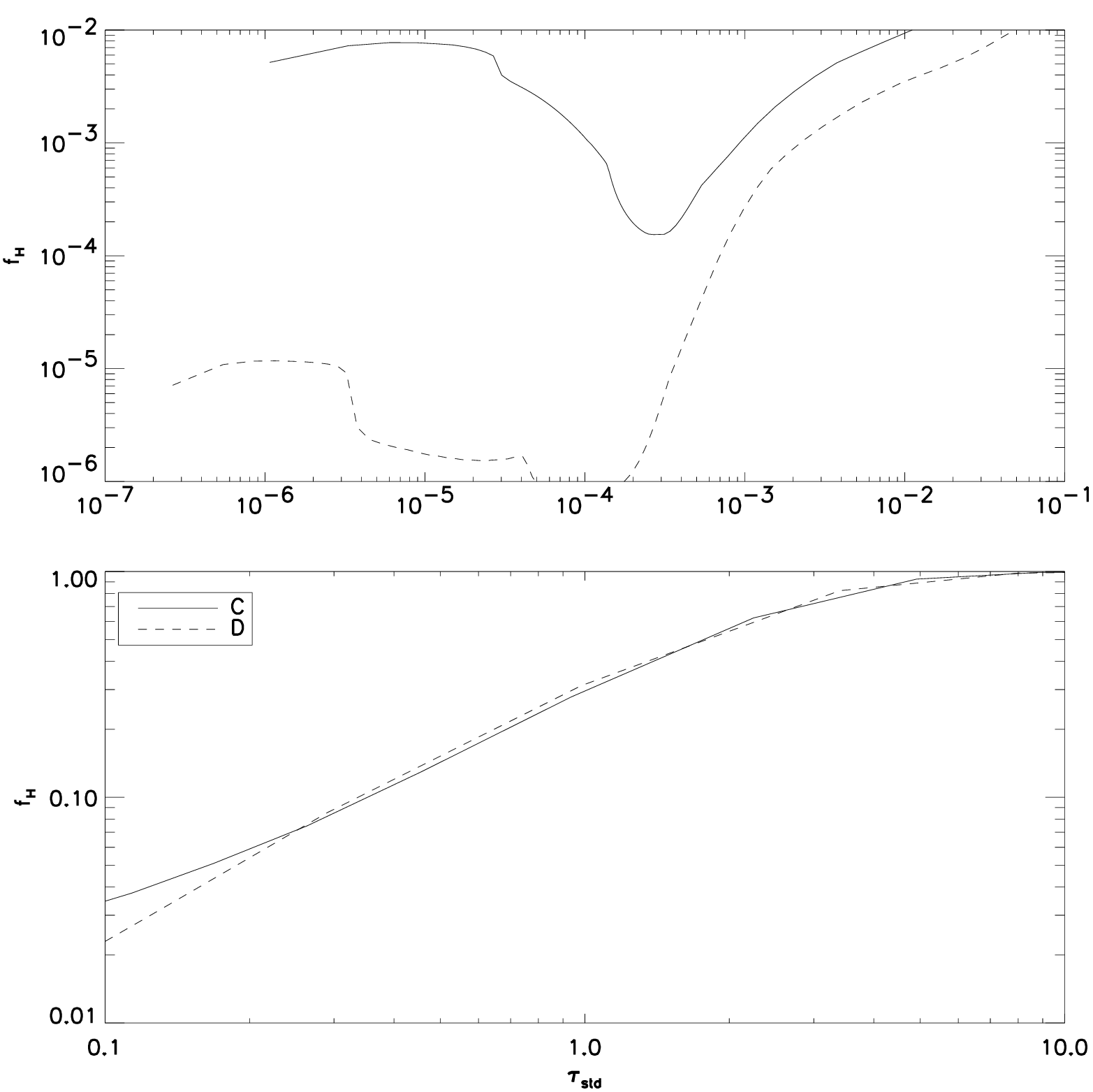}
\caption{Comparison of the hydrogen ionization fraction found using  the 4
  level (C) and 921-level (D) model atoms in a metal-rich environment. 
The upper panel shows the lower optical depth regime while the lower panel
  shows the higher optical depth regime. We define
$\tau_{std}$ as the optical depth corresponding to the continuum
opacity at 500~nm. 
  \label{ionfrac_CD}}
\end{figure*}

We first discuss our results for the models where the composition included metals.
 We treat helium and  metals 
in LTE. Hydrogen is always treated in NLTE.
Recall that these systems
are termed Models C and D  for the 4-level and 921-level
hydrogen atom models, respectively. Figure~\ref{ionfrac_CD} shows
the
hydrogen ionization fraction for Models C and D. There is a
significant change in the hydrogen ionization level, $f_{H}$, between
Models C and D. The quantity $f_{H}$ decreases in the multi-level atom case
in the lower optical depth regime. For $\tau_{std}>0.1$ the
ionization levels among Models C and D are not very different. 
 The quantitative
difference is also tabulated in Table~$\ref{equilibrium_table}$ which shows
the physical parameters for Models C and
D when the 2$\gamma$ process was or was not included.  
 The reduction in $f_{H}$ due to additional
angular momentum sub-states was about a factor of 3 at an optical
depth of about $\tau_{std}
\sim 10^{-4}$ (when the 2$\gamma$ process was included in both the
models). The difference in $f_{H}$ decreases as the optical depth
increases. In metal-rich systems, (Models C and D),
the exclusion of the 2$\gamma$ process did not seem to have any
significant effect for almost all optical depths of interest.

\begin{figure*}
\centering
\includegraphics[width=0.65\textwidth,angle=90]{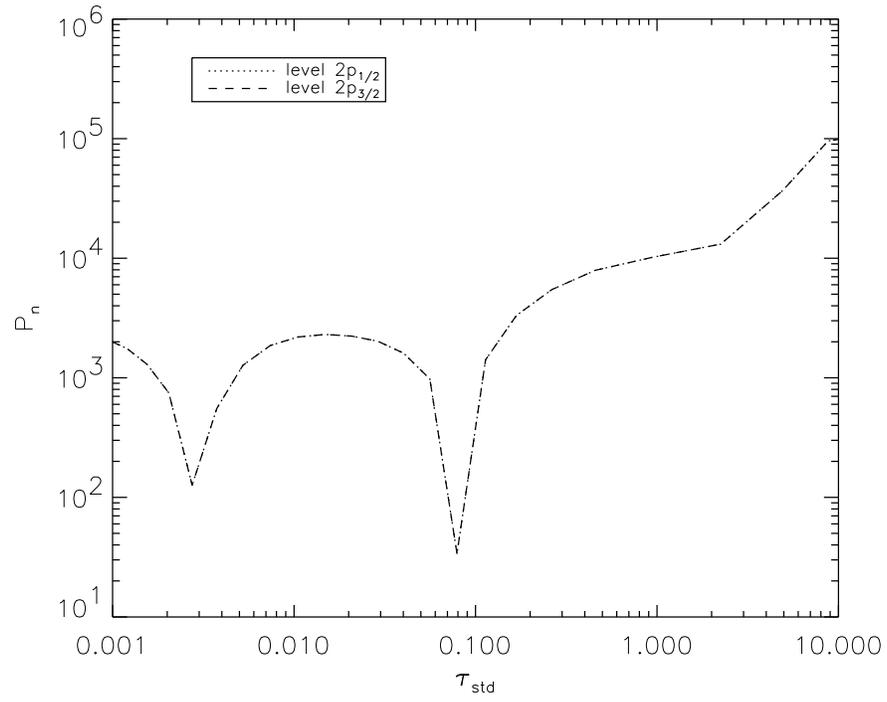}
\caption{Photo-ionization rates of the $2p$ states versus optical depth for  Model C.
\label{photo_C}}
\end{figure*}

\begin{figure*}
\centering
\includegraphics[width=0.65\textwidth,angle=90]{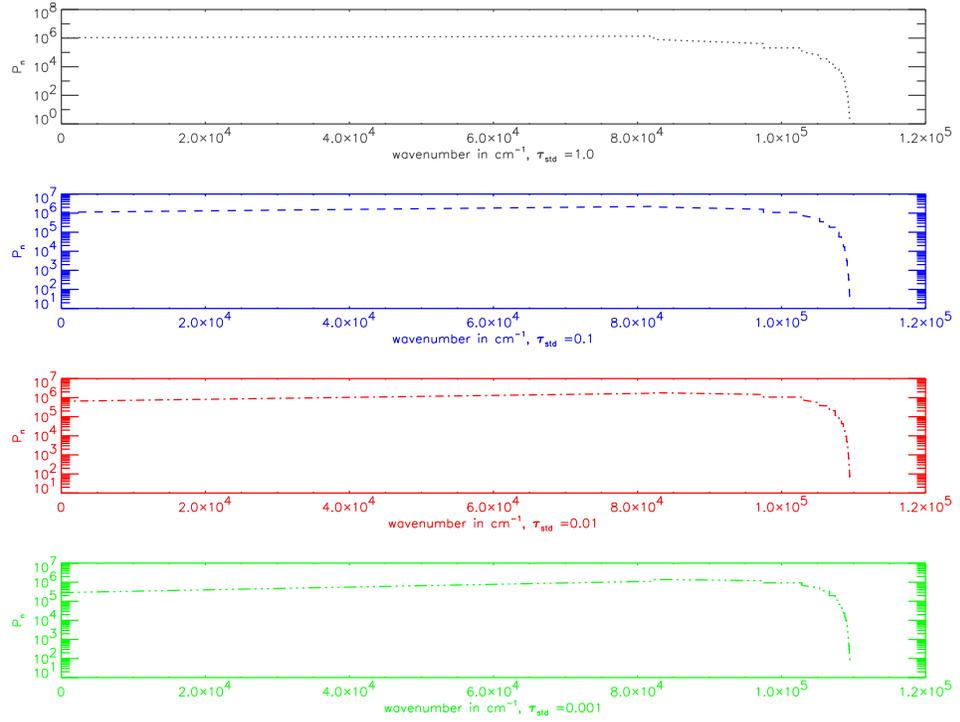}
\caption{Photo-ionization rates versus energy level for Model D at different
  optical depths. Each panel shows  a particular optical depth.
\label{photo_D}}
\end{figure*}

Figure~\ref{photo_C}
displays the net photo-ionization rate from any 
bound state of the hydrogen atom, $P_{n}$, for
the levels 2$p_{1/2}$ and 2$p_{3/2}$ for Model C. The profile of
$P_{n}$ for Model C is not monotonic, but there is an overall
 trend to increase with optical depth, $\tau_{std}$.  This increase
in $P_{n}$ is expected due to the fact that  photo-ionization 
dominates  over recombination at higher optical depths  due to higher
temperatures.  Figure~\ref{photo_D} shows the net
photo-ionization rate as a function of wave number of each energy
level for  Model D.  Different panels show 
different $\tau_{std}$ regimes.
The $P_{n}$ values increase with increasing
$\tau_{std}$ for a given energy level. Also the net
photo-ionization rate does not change significantly with the
change in the energy level of the bound state.
There is a drop in the $P_{n}$
profile at energy levels very close to the continuum. This may be due
to the fact that very high energy bound states are not really
distinguished from the continuum, therefore the \emph{net} photo-ionization
rate falls off.

\begin{figure*}
\centering
\includegraphics[width=0.65\textwidth,angle=90]{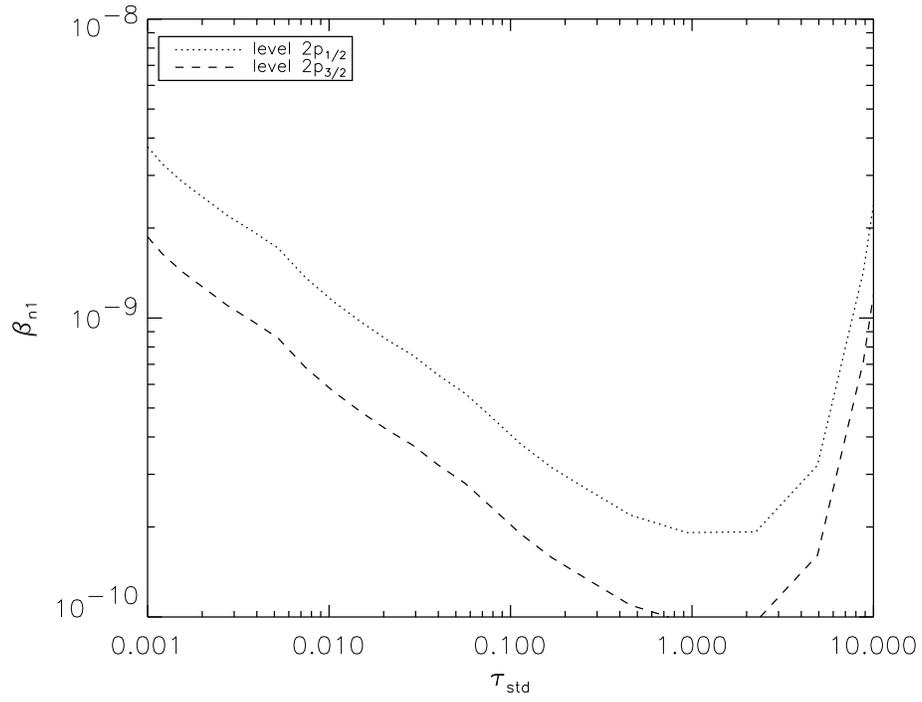}
\caption{Escape probability versus optical depth for the $2p$ states for  Model C.
  \label{esc_C}}
\end{figure*}

\begin{figure*}
\centering
\includegraphics[width=0.65\textwidth,angle=90]{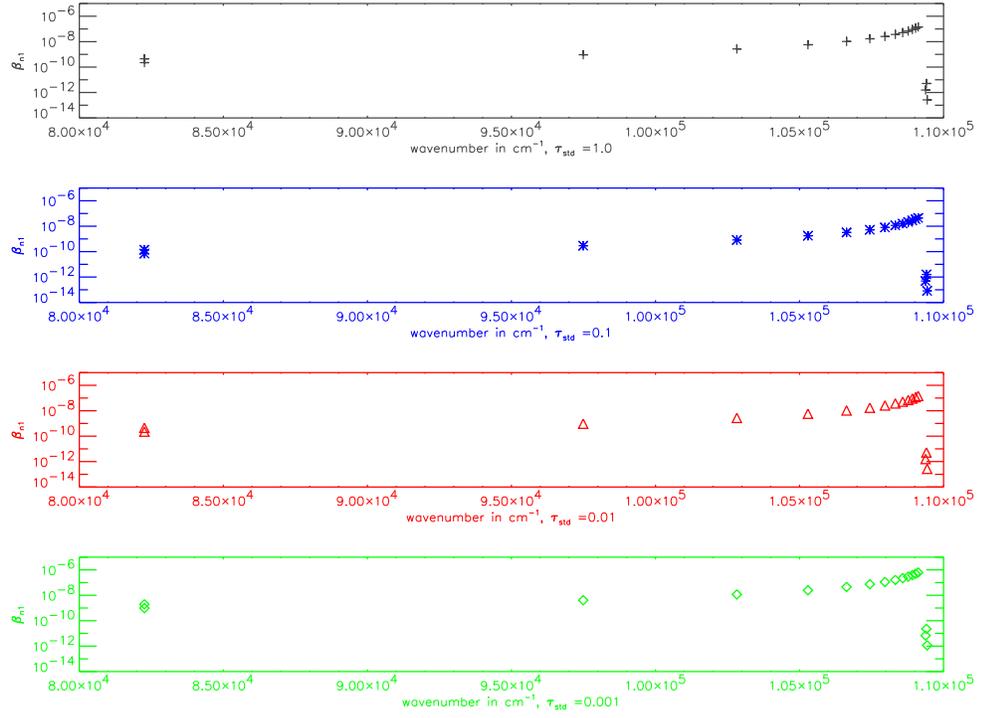}
\caption{Escape probability versus energy levels for  Model D at different optical depths.
Each panel shows  a particular optical depth.
\label{esc_D}}
\end{figure*}

For the calculation of the escape probability (defined as the
probability for the escape of the resonant line connecting the ground
state and higher energy bound state), are shown in Figures~\ref{esc_C}--\ref{esc_D}.
Figure~\ref{esc_C} shows the escape probability for Model C, for 
levels 2$p_{1/2}$ and 2$p_{3/2}$. The escape probability for Model
C decreases with increasing optical depth and also decreases with
decreasing energy of the bound state.  For Model D (Figure
\ref{esc_D}) the escape probability increases with increasing energy
of the bound state.  Also the escape probability decreases very
slowly with the increase in optical depth for Model D, similar to that
in Model
C. The apparent difference between Models C and D is
that the escape probability decreased for a higher energy level in Model  C as opposed to case D where the escape probability
increased with the increased energy of the bound state.  For  Model
C, the levels in question are both $2p$ states and in Model 
D for those two 2$p$ states we also find a
slight reduction in the escape probability with increasing  energy
of the bound state.

The reduction in the escape probability as a
function of optical depth can be explained from
Eq.~\ref{formula_beta}. At higher optical depths (for a given excited 
state), when  $n_{n}$, the population of the excited state increases, this 
results in a decrease in the quantity $\eta_{n1}$. Following the decrease
in $\eta_{n1}$, the escape probability $\beta_{n1}$ decreases. On the
other hand at a given optical depth with increasing energy of
the bound state, the escape probability increases.  At a given
$\tau_{std}$,  for an increase in the bound state energy, $n_{n}$
decreases and $\lambda_{n1}$ decreases, hence the overall effect almost
always is to decrease  $\beta_{n1}$.

\begin{figure*}
\centering
\includegraphics[width=0.65\textwidth,angle=90]{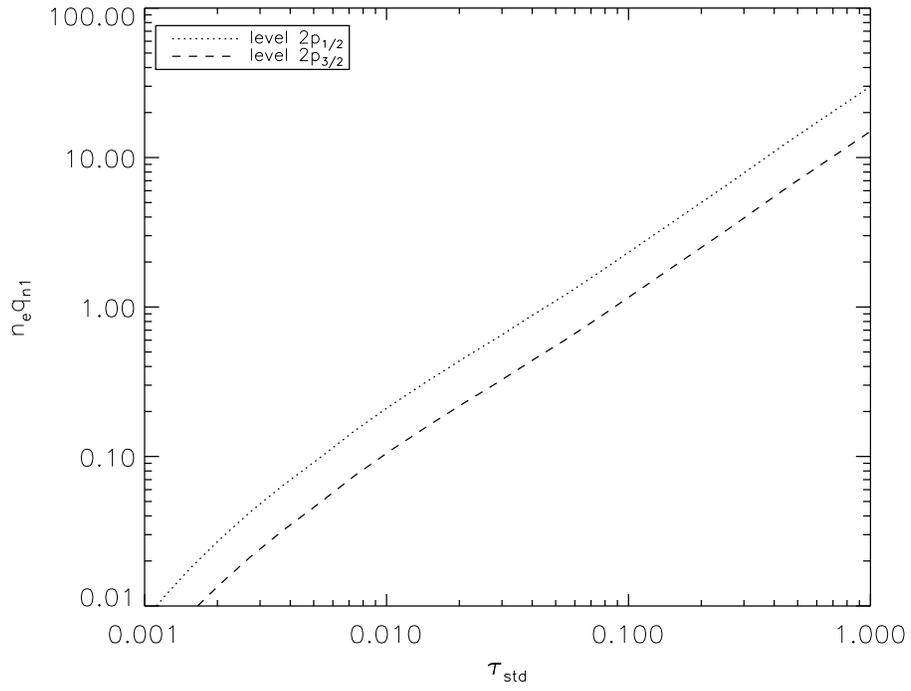}
\caption{Collisional de-excitation coefficients of the $2p$ states
  versus optical depth for  Model C.
\label{coll_C}}
\end{figure*}

Figure \ref{coll_C} shows the collisional de-excitation rate (which is the collisional de-excitation
coefficient $q_{e}$ times the free electron density $n_{e}$) for 
Model C. In this Model there is an increase in the collisional
de-excitation rate with increasing optical depth. For
Model D, the collisional de-excitation rate follows a similar pattern.
The increase 
in the free electron density with increasing $\tau_{std}$ explains the
increased collisional de-excitation rate with $\tau_{std}$. To
summarize, the characteristics of the metal-rich Models C and D, 1) the
ionization fraction decreases for a multi-level hydrogen atom compared
to a 4-level hydrogen atom.
2) The net photo-ionization rate increases
with optical depth for a given bound state. The photo-ionization
rate also does 
not change significantly at a given $\tau_{std}$ as the energy of 
the bound state increases.  3) The escape probability decreases with
increasing  $\tau_{std}$, although this decrease is very small.
The escape probability increases with increasing
energy of the bound state except for the case of the $2p$ states. 4) The
collisional rate increases with increasing optical depth. Also
for the multi-level case the rate is close to the equilibrium value.  5) The
2$\gamma$ process did not seem to have any significant effects in 
Models C and D.

\subsection{Metal-deficient Models}

\begin{figure*}
\centering
\includegraphics[width=0.65\textwidth,angle=0]{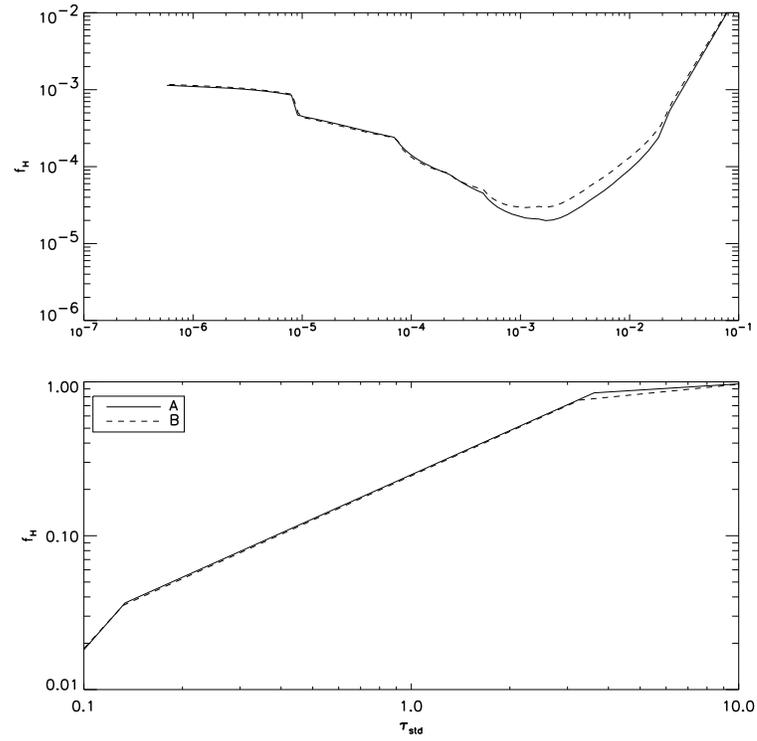}
\caption{Comparison of hydrogen ionization fraction obtained using the 4 level (A)  and the 921-level model (B) hydrogen atom in pure hydrogen
  under radiative equilibrium. The upper panel shows the lower 
optical depth regime while the lower panel shows the higher optical 
depth regime. 
\label{ionfrac_AB_tcor}}
\end{figure*}

In this section we focus on Models A and B. Recall that Models A and B
are models where the atmosphere consists of pure hydrogen. Hydrogen was
treated in NLTE in these models. In Model A, we use a 4-level hydrogen
atom and in Model B, we use a hydrogen atom model with 921 energy
levels.  We also study how the 2$\gamma$ process affects these 
systems when the environment is
metal-free. Figure~\ref{ionfrac_AB_tcor} displays the ionization
fraction for Models A and B. Interestingly $f_{H}$ does not change
significantly between Models A and B. This is quite different when
compared to Models C and D.  Table~\ref{equilibrium_table} shows the
changes due to the 2$\gamma$ process. For Model A, at low optical
depths ($\tau_{std} \sim 10^{-4}$) turning on the 2$\gamma$ process
produces about a 20\% change in the free electron density, but for
$\tau_{std} \ga 10^{-3}$ no significant change is seen.  For Model B,
there is a significant reduction in the free electron density due to
the inclusion of the 2$\gamma$ process. The electron density is 5
times higher with the inclusion of the 2$\gamma$ process in case B, at
low optical depths, $\tau_{std} \sim 10^{-4}$. For higher optical
depths, there is no significant effect due to the 2$\gamma$ process.
The decrease in the free electron density due to the 2$\gamma$ process
is expected since the upward process is suppressed by a factor of
1/137.  The change is especially noticeable in Model B.  It is at a fairly
low optical depth where the change in the rates observed
but this is also the optical depth where the optical depth of the Balmer line is
high \citep{soma09}. This wavelength regime is similar to what is described in
\citet{soma09}.

\begin{figure*}
\centering
\includegraphics[width=0.65\textwidth,angle=90]{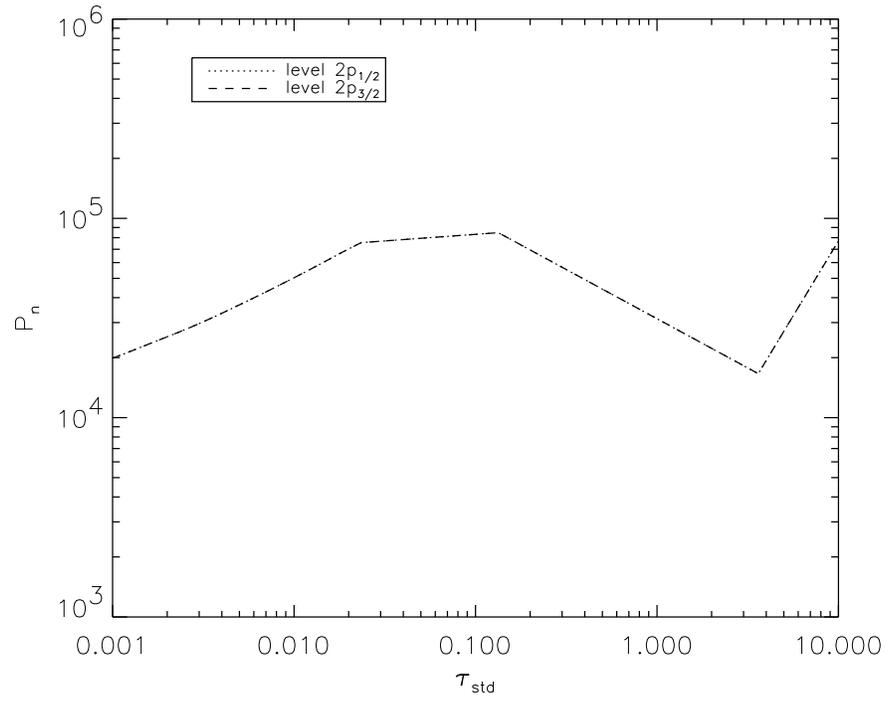}
\caption{Photo-ionization rate versus optical depth for $2p$ states of  Model A
  under radiative equilibrium. The last point at $\tau_{std} = 10$ is
  a numerical glitch caused by poor spatial resolution. The rate
  should continue to drop with depth.
\label{photo_A}}
\end{figure*}

\begin{figure*}
\centering
\includegraphics[width=0.65\textwidth,angle=90]{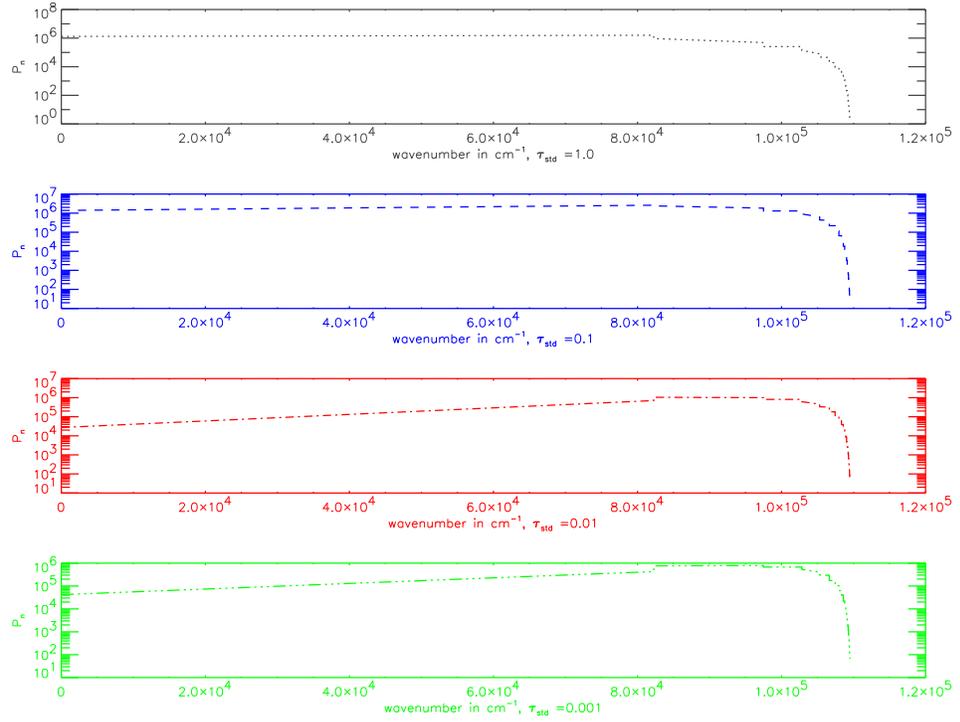}
\caption{Photo-ionization rates versus energy level for Model B in
  radiative equilibrium.
  Each panel shows  a particular optical depth.
\label{photo_B}}
\end{figure*}

\begin{figure*}
\centering
\includegraphics[width=0.65\textwidth,angle=90]{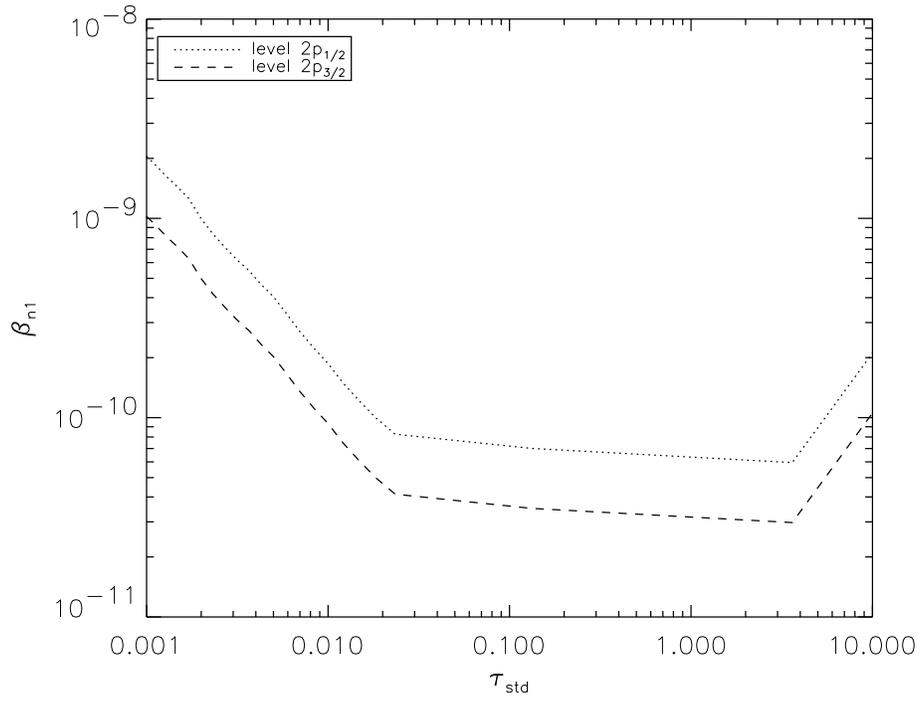}
\caption{Escape probability versus optical depth for the $2p$ states of  Model A in
  radiative equilibrium. 
\label{esc_A}}
\end{figure*}

\begin{figure*}
\centering
\includegraphics[width=0.65\textwidth,angle=90]{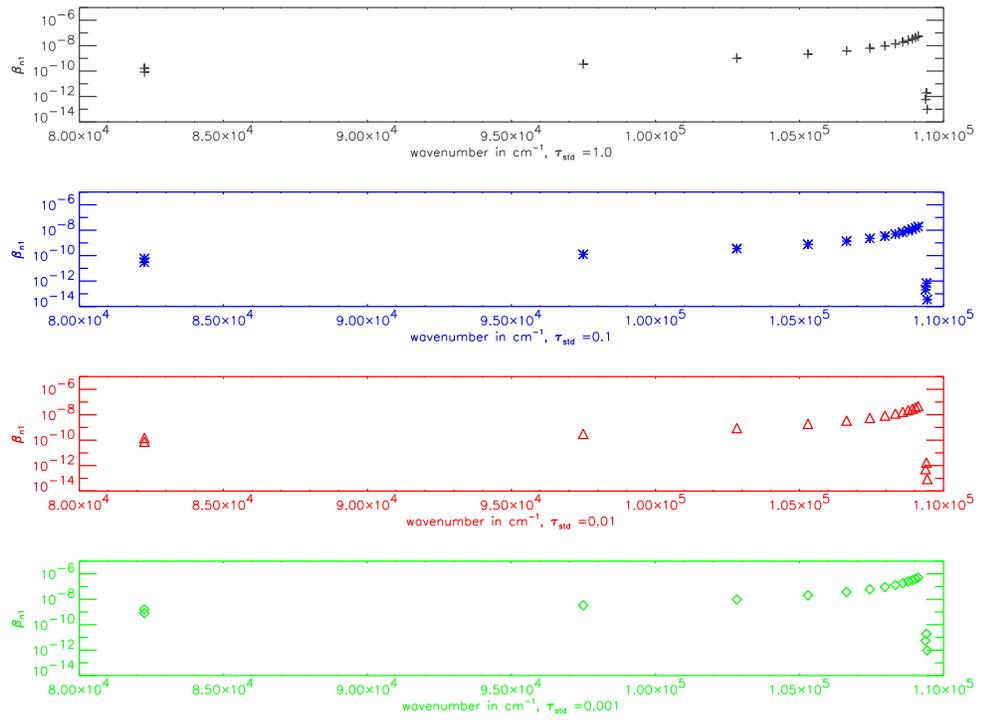}
\caption{Escape probability versus energy level for  Model B in radiative
equilibrium. Each panel refers to a specific optical depth.
\label{esc_B}}
\end{figure*}

The net photo-ionization rate ($P_{n}$) does not
change significantly between different bound states in both Models A
and B (see Figures~\ref{photo_A} and \ref{photo_B}). Although in the
multi-level case (case B), $P_{n}$ increases by almost a factor of 10
from the lowest energy bound state to the higher energy bound
states (except for the states very close to the continuum) at low
optical depths ($\tau_{std} <0.01$). Figures~\ref{esc_A}
and \ref{esc_B} 
show the escape probability for  Models A and B
respectively. The trend is very similar to that of Models C and D.
 The collisional de-excitation rate  increases 
with increasing optical depths in both the Models C and D

To summarize: our findings on Models A and B (pure hydrogen models)
1) For most optical depth regimes, the basic trend in
the rates is similar to Models C and D.
2)The 2$\gamma$ process seems to have a significant effect in both
Model A and B (at lower $\tau_{std}$). The effect is much larger
for Model B. This effect was not seen in Models C and D.

We observe the following, for almost all optical depths,
\begin{eqnarray}
P_{n}(A) &>& P_{n}(C) ;\nonumber \\
\beta_{n1}(A) &<& \beta_{n1}(C) ;\nonumber \\
n_{e}q_{n1}(A) &<& n_{e}q_{n1}(C). \nonumber
\end{eqnarray}
For Models B and D, 
\begin{eqnarray}
P_{n}(B) &<& P_{n}(D)\ \ (\tau < 0.01) ; \nonumber \\
\beta_{n1}(B) &\approx& \beta_{n1}(D) ;\nonumber \\
n_{e}q_{n1}(B) &<& n_{e}q_{n1}(D)\ \ (\tau < 0.01). \nonumber
\end{eqnarray}
The difference in the profiles of $P_{n}$, $\beta_{n1}$ and 
$n_{e}q_{n1}$ between Model A (or B) and Model C (or D) can be
attributed to their different temperature profiles.  

\begin{table*}
\centering
\caption[modeltab]{\label{equilibrium_table}
Physical characteristics from four different models, under radiative equilibrium}
\begin{tabular}{c rrrrr}
\hline&\\
Model Name & Type &$\tau$ & $T$ (K) & $N_{e}$ (cm$^{-3}$) & $\mu$ \\ 
\hline & \\
\  & \ &0.00016  & 2820.75      &105236. &1.01 \\
\ A &4 level &0.02343  &4011.17  &9.999E+07 &1.01 \\
\ & \ & 3.61 & 7564.24  & 2.351E+11 & 0.54 \\ 
\  &\ &1104.90  &27565.10 & 3.903E+12 &0.50\\
\hline
\   &\  &0.00016  &2820.61    &  121450. &1.01 \\
\ A' & 4 level, no 2$\gamma$ process &0.02345    & 4011.18  & 9.999E+07  &  1.01 \\
\ & \ & 3.61 & 7564.29  & 2.351E+11 & 0.54 \\
\   &\  &1106.10     & 26840.2  & 3.9035E+12 & 0.50\\
\hline
\   &\ &0.00016   & 2801.08 & 3.946E+5 & 1.01\\
\ B  &921 level &0.0272   & 4547.80  &2.981E+8 & 1.01\\
\   & \ & 1.794 & 6779.06 & 1.087E+11 & 0.57 \\
\   &\  &1181.       & 20462.42 & 3.893E+12 & 0.50\\
\hline
\  & \ &0.000157   & 2826.28    & 6.709E+4 &1.01\\
\  B' & 921 level, no 2$\gamma$ process &0.02705 & 4547.69 & 2.983E+08 &1.01\\
\   & \ & 1.794 & 6778.83 & 1.0864E+11 & 0.57 \\
\   &\ &1181    &  20462.42  &3.801E+12& 0.50\\
\hline
\ & \ &0.00023  &  3564.79    &  271828. &1.26\\
\  C & 4 level, metal-rich environment&4.91280     & 7584.85  & 1.260E+11& 0.68\\
\  &\ &867.030    & 26365.40 & 3.162E+12& 0.625\\
\hline
\ &\  &0.00023  &3564.78  &    271808. & 1.26\\
\  C' &4 level, metal rich environment &4.91270 & 7584.84 & 1.260E+11& 0.68\\
\ &no 2$\gamma$ process & 867.900   &  26243.10  &3.160E+12 &0.625\\
\hline
\ &\ &0.00011 & 3099.54     & 84640.4 &1.26\\
\  D  & 921 level, metal rich environment &0.98338  &6337.20  &4.254E+10 &0.98\\
\   &\ &863.240  &  26109.80  &3.156E+12 &0.63\\
\hline
\  &\  & 0.00011 &  3099.46     & 84638.0 &1.26\\
\ D' & 921 level, metal rich environment &0.98373 &  6337.44 & 4.256E+10 &0.98\\
\  & no 2$\gamma$ process &863.080   &  26113.50  &3.156E+12 & 0.63\\
\hline
\end{tabular}
\end{table*}

\subsection{Metal-deficient case without temperature corrections}

In our previous Models A--D, we allowed the system to reach radiative
equilibrium for each assumption. It is useful to hold the temperature
structure fixed and just examine the changes that are due to the
variation in the compositions and model atoms.
In the fixed temperature structure case we do not find a
large difference as compared to the radiative equilibrium case.
The differences, compared to the
radiative equilibrium case occurs in the optical
depth region $10^{-3}<\tau <10^{-2}$. There is a slight suppression in
the ionization fraction for Model B compared to  Model A. The other
rates are nearly identical.

\section{Discussion }\label{sec:discuss}

The primary motivation for this paper was to investigate how the
multi-level structure of the hydrogen model atom  affects the hydrogen
ionization fraction as well as the photo-ionization, escape, and
collisional de-excitation rates. We also investigated if the
the metals in the environment or if the  2$\gamma$ process
plays a significant role in the recombination process of the system. 
 We find that a multi-level hydrogen atom structure, or
in other words  a more complete set of energy levels
to be important in determining the ionization profile of the
system. Recently, \citet{grin_hirata10} studied the effects of
including a more complete set of angular momentum substates in the
study of cosmological recombination and also found that including
more substates to be important. The need to have a more
complete set of bound 
states in the hydrogen atom is found to be more important in a metal-rich
environment. This is because of the fact that 
recombination is more effective in a multi-level framework. When there
is a source of additional free electrons (such as metals), the
suppression in $f_{H}$ is larger than in pure hydrogen
models. This is merely due to the fact that the larger free electron
density drives recombination,
thus it is very important to use multi-level hydrogen atoms
especially in realistic environments with solar compositions.

The enhancement of recombination is indicated by the 
increasing escape probability for the higher bound states (for both
Models B and D). This is also reinforced by the almost constant
photo-ionization rate over different energy bound states.

\begin{figure*}
\centering
\includegraphics[width=0.65\textwidth,angle=90]{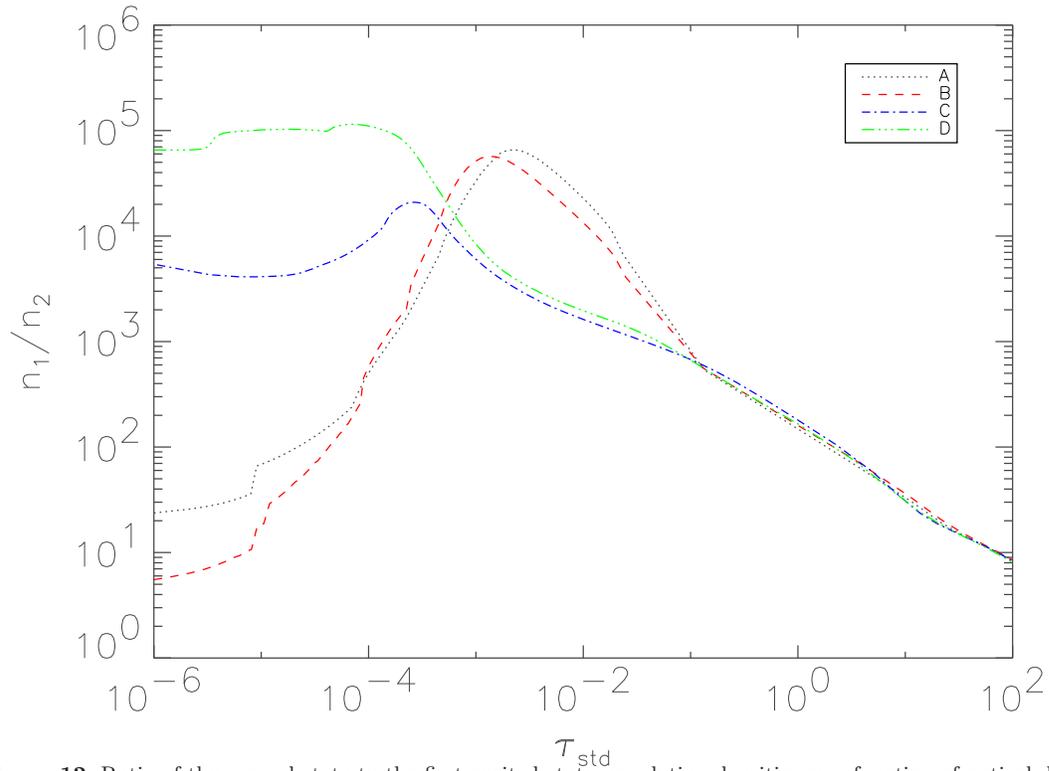}
\caption{Ratio of the ground state to the  first excited state
  population densities as a function of optical depth.
\label{bi_1}}
\end{figure*}

The importance of the 2$\gamma$ process is seen in the 4-level
pure hydrogen case (Model A) as only a small reduction in the free electron
density. In Model B, the reduction in the free electron density
due to the inclusion of the 2$\gamma$ process is much larger. There
was not any significant effect due to the 2$\gamma$ process in the
solar composition Models. 
The 
larger change in Model B (at lower $\tau_{std}$) can be
explained from Figure \ref{bi_1}. At low optical depths the
relative population of the ground state to the first excited state is
displayed. For the metal-free Models (A and B) the ratio
$\frac{n_{1}}{n_{2}}$ is not very high and about 4.0 for
Model B. Thus in Model B, the first excited state
population is not significantly lower than the ground state
population. The relative first excited state population is much lower
in Models C and D (where the ratio is around $10^{4}-10^{5}$
at similar $\tau_{std}$).  The high level population of the first
excited state (for Model B) increases the importance of the 
 2$\gamma$ process, due to the higher number of available electrons that can recombine into the ground state and the upward transition probability for this non-resonant process being low, thus the recombination obtained is more effective.
We find that it is essential to incorporate both multi-level atoms and the 2$\gamma$ process.
Our primary purpose was to do a simple calculation to emphasize
the important factors in determining the recombination 
of hydrogen at typical free electron densities found in Type II
supernova atmospheres.
Our results show that it is important to 
incorporate 2$\gamma$ transitions and multiple angular momentum
sub-states at low optical depths where the free electron density is
small (typically $\sim 10^{5}$~cm$^{-3}$) and is also typical of the free electron density during the epoch of cosmological recombination
\citep{Peebles}. In H II regions the electron density is much lower 
making the hydrogen recombination time scale rise. Therefore accurate treatment of hydrogen recombination is important in all of these scenarios. In the future we plan to investigate hydrogen recombination in the context of cosmological recombination epoch.

\section*{Acknowledgments} 
We thank the anonymous referee for helpful comments which
significantly improved the presentation of this work.
This work was supported in part by NSF grant AST-0707704, Department
of Energy Award Number DE-FG02-07ER41517, and by SFB grant 676 from
the DFG.  This research used resources of the National Energy Research
Scientific Computing Center (NERSC), which is supported by the Office
of Science of the U.S.  Department of Energy under Contract No.
DE-AC02-05CH11231; and the H\"ochstleistungs Rechenzentrum Nord
(HLRN).  We thank both these institutions for a generous allocation of
computer time.

\bibliography{apj-jour,refs,bib_recom,baron,sn1bc,sn1a,sn87a,snii,stars,rte,cosmology,gals,agn,atomdata,crossrefs}

\begin{thebibliography}{}

\bibitem[\protect\citeauthoryear{{Chluba} \& {Sunyaev}}{{Chluba} \&
  {Sunyaev}}{2006}]{CS06}
{Chluba} J.,  {Sunyaev} R.~A.,  2006, \aap, 446, 39

\bibitem[\protect\citeauthoryear{{Chluba} \& {Sunyaev}}{{Chluba} \&
  {Sunyaev}}{2008}]{CS08b}
{Chluba} J.,  {Sunyaev} R.~A.,  2008, \aap, 480, 629

\bibitem[\protect\citeauthoryear{De, Baron \& Hauschildt}{De
  et~al.}{2009}]{soma09}
De S.,  Baron E.,    Hauschildt P.~H.,  2009, MNRAS, 401, 2081

\bibitem[\protect\citeauthoryear{{Dessart} \& {Hillier}}{{Dessart} \&
  {Hillier}}{2008}]{dessart08}
{Dessart} L.,  {Hillier} D.~J.,  2008, \mnras, 383, 57

\bibitem[\protect\citeauthoryear{{Dubrovich} \& {Grachev}}{{Dubrovich} \&
  {Grachev}}{2005}]{DG05}
{Dubrovich} V.~K.,  {Grachev} S.~I.,  2005, Astronomy Letters, 31, 359

\bibitem[\protect\citeauthoryear{{Fendt}, {Chluba}, {Rubi{\~n}o-Mart{\'{\i}}n}
  \& {Wandelt}}{{Fendt} et~al.}{2009}]{FCRW09}
{Fendt} W.~A.,  {Chluba} J.,  {Rubi{\~n}o-Mart{\'{\i}}n} J.~A.,    {Wandelt}
  B.~D.,  2009, \apjs, 181, 627

\bibitem[\protect\citeauthoryear{{Grin} \& {Hirata}}{{Grin} \&
  {Hirata}}{2010}]{grin_hirata10}
{Grin} D.,  {Hirata} C.~M.,  2010, Phys Rev D, 81, 083005

\bibitem[\protect\citeauthoryear{Hauschildt \& Baron}{Hauschildt \&
  Baron}{1999}]{hbjcam99}
Hauschildt P.~H.,  Baron E.,  1999, J. Comp. Applied Math., 109, 41

\bibitem[\protect\citeauthoryear{Hauschildt \& Baron}{Hauschildt \&
  Baron}{2004}]{hbmathgesel04}
Hauschildt P.~H.,  Baron E.,  2004, Mitteilungen der Mathematischen
  Gesellschaft in Hamburg, 24, 1

\bibitem[\protect\citeauthoryear{{Hirata}}{{Hirata}}{2008}]{Hirata08}
{Hirata} C.~M.,  2008, \prd, 78, 023001

\bibitem[\protect\citeauthoryear{{Hirata} \& {Forbes}}{{Hirata} \&
  {Forbes}}{2009}]{HF09}
{Hirata} C.~M.,  {Forbes} J.,  2009, \prd, 80, 023001

\bibitem[\protect\citeauthoryear{{Hirata} \& {Switzer}}{{Hirata} \&
  {Switzer}}{2008}]{HS08}
{Hirata} C.~M.,  {Switzer} E.~R.,  2008, \prd, 77, 083007

\bibitem[\protect\citeauthoryear{{Karshenboim} \& {Ivanov}}{{Karshenboim} \&
  {Ivanov}}{2008}]{KI08}
{Karshenboim} S.~G.,  {Ivanov} V.~G.,  2008, Astronomy Letters, 34, 289

\bibitem[\protect\citeauthoryear{{Kholupenko}, {Ivanchik} \&
  {Varshalovich}}{{Kholupenko} et~al.}{2007}]{KIV07}
{Kholupenko} E.~E.,  {Ivanchik} A.~V.,    {Varshalovich} D.~A.,  2007, \mnras,
  378, L39

\bibitem[\protect\citeauthoryear{{Kurucz}}{{Kurucz}}{1995}]{Kurucz}
{Kurucz} R.~L.,  1995, Highlights of Astronomy, 10, 579

\bibitem[\protect\citeauthoryear{Mihalas}{Mihalas}{1978}]{mihalas78sa}
Mihalas D.,  1978, Stellar Atmospheres.
W. H. Freeman, New York

\bibitem[\protect\citeauthoryear{{Peebles}}{{Peebles}}{1968}]{Peebles}
{Peebles} P.~J.~E.,  1968, \apj, 153, 1

\bibitem[\protect\citeauthoryear{{Rubi{\~n}o-Mart{\'{\i}}n}, {Chluba} \&
  {Sunyaev}}{{Rubi{\~n}o-Mart{\'{\i}}n} et~al.}{2008}]{RCS08}
{Rubi{\~n}o-Mart{\'{\i}}n} J.~A.,  {Chluba} J.,    {Sunyaev} R.~A.,  2008,
  \aap, 485, 377

\bibitem[\protect\citeauthoryear{{Seager}, {Sasselov} \& {Scott}}{{Seager}
  et~al.}{1999}]{SSS99}
{Seager} S.,  {Sasselov} D.~D.,    {Scott} D.,  1999, \apjl, 523, L1

\bibitem[\protect\citeauthoryear{{Seager}, {Sasselov} \& {Scott}}{{Seager}
  et~al.}{2000}]{SSS00}
{Seager} S.,  {Sasselov} D.~D.,    {Scott} D.,  2000, \apjs, 128, 407

\bibitem[\protect\citeauthoryear{{Switzer} \& {Hirata}}{{Switzer} \&
  {Hirata}}{2008a}]{SH08a}
{Switzer} E.~R.,  {Hirata} C.~M.,  2008a, \prd, 77, 083006

\bibitem[\protect\citeauthoryear{{Switzer} \& {Hirata}}{{Switzer} \&
  {Hirata}}{2008b}]{SH08b}
{Switzer} E.~R.,  {Hirata} C.~M.,  2008b, \prd, 77, 083008

\bibitem[\protect\citeauthoryear{{Utrobin} \& {Chugai}}{{Utrobin} \&
  {Chugai}}{2005}]{Chugai}
{Utrobin} V.~P.,  {Chugai} N.~N.,  2005, \aap, 441, 271

\bibitem[\protect\citeauthoryear{{Wong}, {Moss} \& {Scott}}{{Wong}
  et~al.}{2008}]{WMS08}
{Wong} W.~Y.,  {Moss} A.,    {Scott} D.,  2008, \mnras, 386, 1023

\bibitem[\protect\citeauthoryear{{Wong} \& {Scott}}{{Wong} \&
  {Scott}}{2007}]{WS07}
{Wong} W.~Y.,  {Scott} D.,  2007, \mnras, 375, 1441

\bibitem[\protect\citeauthoryear{{Zeldovich}, {Kurt} \& {Syunyaev}}{{Zeldovich}
  et~al.}{1969}]{Zeldovich}
{Zeldovich} Y.~B.,  {Kurt} V.~G.,    {Syunyaev} R.~A.,  1969, Soviet Journal of
  Experimental and Theoretical Physics, 28, 146

\end{thebibliography}

\clearpage



\end{document}